\documentstyle{article}

\title{Avalanche consumption and the stationary regions of the density
profile around the droplet in the theory of condensation}
\author{V.Kurasov}

\date{Victor.Kurasov@pobox.spbu.ru}

\begin{document}
\maketitle

The kinetic theory of the condensation was significantly developed during
the last years. Creation of the theoretical description  of the condensation
in frames of the mean field approach started the based quantitative
description
of this process. The inclusion of the profiles of the vapor density around
every  droplet leads to the new view on the condensation. As it is shown
in  \cite{PhysA}  one can apply the formalism of the Green function to
see the form of the density profile around the droplet. Then the natural
idea to see where the mean field approach can be applied can be considered.

The Green function of the diffusion equation has the known form
$$
G \sim \exp(- \frac{r^2}{4Dt} ) / (Dt)^{3/2}
$$
where $r$ is the distance from the droplet, $t$ is the characteristic
time, $D$ is the diffusion coefficient.

>From the first point of view the characteristic distance of the stationary
relaxed profile is given from the argument of the exponent in the expression
for the Green function
$$
r_{rel} \sim  \sqrt{D t}
$$
The problem is to decide what time one has to put in the last formula.
Naturally it is reasonable to put the characteristic  duration of the
period of the nucleation (or the formation of the main quantity of the
supercritical embryos).

On the other hand the essential feature of the condensation process which
allowed to formulate the property of the universality of the droplets
size spectrum is the avalanche character of the vapor consumption by the
droplet. This lies in contradiction with the last estimate.

The aim of this activity is resolve this problem and to give more specified
estimate which allows to define concrete region of the stationarity of
the vapor density profile around the droplet.

We shall start from the expression \cite{PhysA} for
the difference of the supersaturation
$\zeta$ from the ideal supersaturation $Phi$
$$
\frac{\Phi - \zeta}{\Phi} =
\sqrt{\frac{2}{\pi}} (\frac{v_l}{v_v})^{1/2} f(\beta)
$$
where
$v_v$ and $v_l$ are the volumes for the molecule in the vapor phase and
in the liquid phase and $f(\beta)$ is the function of the argument
$$
\beta = \frac{r}{\sqrt{4Dt}}
$$
For the concrete form of $f(\beta)$ one can get
$$
f(\beta) =
\int_{\beta}^{\infty}
(\frac{1}{\beta^2} - \frac{1}{x^2} )^{1/2}
\exp(-x^2) dx
$$

In \cite{eprint2} it was shown that the mean field approach can be applied
when parameter
$$
\sigma = \Gamma^2 \frac{v_l}{v_v}
$$
is small in comparison with unity\footnote{The parameter $\Gamma$ is defined
in \cite{PhysA}.} .
In terms of $\sigma$ one can rewrite the required expression as
$$
\frac{\Phi - \zeta}{\Phi} =
\sqrt{{2}{\pi}} \frac{\sigma^{1/2}}{\Gamma} f(\beta)
$$

Obtain the exact expression for $f(\beta)$. One can rewrite the mentioned
expression as
$$
f(\beta) =
2 \beta^2 \Gamma(3/2)
exp(-\beta^2) \Psi(\frac{3}{2}, \frac{3}{2}; \beta^2)
$$
Here
$\Gamma(z)$ is the Gamma-function, $\Psi(x,y;z)$ is the confluent
hypergeometric
function.

The following asymptote for the great values of $\beta$ can be gotten
$$
f(\beta) =
\exp(-\beta^2) \frac{1}{2\beta^3} \int_{0}^{\infty}  y^{1/2} \exp(-y) dy
 \sim \frac{\exp(-\beta^2)}{ \beta^3}
$$

For small values of $\beta$ one can get
$$
f(\beta) =
\frac{\sqrt{\pi}}{2} \frac{1}{\beta} \equiv f_{as}
$$
This asymptote corresponds to the stationary solution for the
density profile
$$
\frac{\Phi - \zeta}{\Phi}  \sim \sigma^{1/2} \frac{1}{\Gamma \beta}
$$

The characteristic value of the error in the relative deviation of $f$
from $f_{as}$ have to attain the same value as the error of the calculation
of the iterations,  i.e. somewhere about 15 percents. Concrete calculations
show
such a deviation is already attained at $\beta \sim 0.05$. It means
that the value of the parameter $sigma$ have to be less than $1/400$.
It means that  the theory based on the stationary profile around the droplet
can be applied only when $\sigma < 1/400$. So, the region of applicability
becomes rather small and one has really use the formalism of the Green
function described in \cite{PhysA}.

It is the avalanche character of the droplets growth which leads to this
great variation of the required value of the parameter $\sigma$ necessary
for the  validity of the mean field approach. Note that the
complete avalanche
character would lead to the absence of the stationary region. We see that
 moderate avalanche character leads to the fact that the stationary
(smaller
than one can assume from the first point of view) region nevertheless
exists. This leads to the applicability of the stationary approximation
for the droplets growth. This fact is well known and based in many papers.

\end{document}